# Double-slit photoelectron interference in strong-field ionization of the neon dimer

Maksim Kunitski[1*], Nicolas Eicke[2], Pia Huber[1], Jonas Köhler[1], Stefan Zeller[1], Jörg Voigtsberger[1], Nikolai Schlott[1], Kevin Henrichs[1], Hendrik Sann[1], Florian Trinter[1], Lothar Ph. H. Schmidt[1], Anton Kalinin[3], Markus Schöffler[1], Till Jahnke[1], Manfred Lein[2] and Reinhard Dörner[1*].

[1]Institut für Kernphysik, Goethe-Universität Frankfurt am Main, Max-von-Laue-Straße 1, 60438 Frankfurt am Main, Germany

[2]Institut für Theoretische Physik, Leibniz Universität Hannover, Appelstraße 2, 30167 Hannover, Germany

[3]GSI-Helmholtz Center for Heavy Ion Research, Planckstraße 1, 64291 Darmstadt, Germany

*Correspondence to: kunitski@atom.uni-frankfurt.de, doerner@atom.uni-frankfurt.de

**Wave-particle duality is an inherent peculiarity of the quantum world. The double-slit experiment has been frequently used for understanding different aspects of this fundamental concept. The occurrence of interference rests on the lack of which-way information and on the absence of decoherence mechanisms, which could scramble the wave fronts. In this letter, we report on the observation of two-center interference in the molecular frame photoelectron momentum distribution upon ionization of the neon dimer by a strong laser field. Postselection of ions, which were measured in coincidence with electrons, allowed choosing the symmetry of the continuum electronic wave function, leading to observation of both, *gerade* and *ungerade*, types of interference.**

Wave-like behaviour of microscopic particles, e.g. interference, and, in general, the particle-wave duality is "the mystery", as stated by Feynman[1], ''which is impossible, absolutely impossible, to explain in any classical way, and which has in it the heart of quantum mechanics''. The double-slit experiment, which was originally conducted by Young in the early 1800s to prove the wave nature of light, has been widely utilized to learn about different aspects of this "mystery".

In the 1960s it was realized that the double-slit experiment can be performed at the molecular level by exploiting two sites of a diatomic molecule as coherent electron emitters[2]. Liberation of an electron from such a system – for instance, upon single photon absorption – will result in interference of two partial waves, spatially separated by the bond length[3–14]. Recent theoretical works[15–17] have suggested that this concept can be transferred to strong-field ionization and thus ultrafast time scales. Reasons possibly contradicting this generalization are that strong-field ionization is often described by tunneling ionization, where the tunnel exit and, thus, the birth position of the electron is not at the atomic centers but at distance comparable to the extent of the molecule. Furthermore, the electrons are accelerated in the laser field and reach their final momentum only at an even larger distance. Thus, it might not be obvious which momentum is relevant for the interference. The presence of the Coulomb field during the acceleration by the laser field causes another potential problem for double-slit interference as it leads, in particular for linearly polarized light, to massive deformation of the phase front. Experimental support for double-slit type interference in the strong-field context are suppression of the ionization efficiency of the oxygen molecule with respect to xenon [18] due to destructive two-center interference. In a pioneering experiment, differences in the photoelectron spectra of non-aligned argon dimers and argon atoms were found and attributed to double-slit interference[19]. More recent work however contradicted this conclusion[20] and attributed similar differences to post-collision interaction.

Here we report on an experiment that clearly shows the full interference fringes from two-center interference in strong-field ionization.

The most natural observable to search for double-slit interference is the momentum component of the interfering particle parallel to the distance vector between the two slits. Superposition of two spherical waves with momentum $\vec{k}$ emerging from two centers separated by $\vec{R}$: $\psi_{1,2} = A\, e^{i(\vec{k}(\vec{r}\, \pm \frac{\vec{R}}{2}) + \phi_{1,2})}$ leads to a probability distribution given by

$$P \sim cos^2(\vec{k} \cdot \frac{\vec{R}}{2} + \frac{\Delta\phi}{2}) \qquad (1)$$

where $\Delta\phi = \phi_1 - \phi_2$ is the initial phase difference between two waves. For the optical double-slit both waves are in phase ($\Delta\phi=0$) and one typically expresses interference in terms of wavelength $\lambda = 2\pi/|\vec{k}|$ and the angle $\vartheta$ with respect to the normal to $\vec{R}$ as $cos^2(\frac{\pi R \sin\vartheta}{\lambda})$.

The experimental challenge, thus, is to measure the projection of the electron momentum onto the molecular axis $\vec{R}$. As the molecules in a gas-phase sample are randomly oriented, this requires measuring the molecular axis for each ionization event in coincidence with the electron momentum. This is possible for Neon dimers, which after single ionization dissociate rapidly along the bond axis into $Ne^0$ and $Ne^+$, thus detection of the direction of the $Ne^+$ in coincidence with the electron allows obtaining the momentum component of the electron along the bond axis.

In the experiment $Ne_2$ was ionized by a 40 fs (FWHM in intensity) 780 nm laser field with intensities of $7.3*10^{14}$ W/cm$^2$ (Keldysh parameter $\gamma=0.72$) in case of circular polarization and $1.2*10^{15}$ W/cm$^2$ ($\gamma=0.4$) for linear polarization. The charged products after ionization were detected in coincidence by COLTRIMS[21]. Here we only consider the single-ionization process that leads to breakup of the dimer into a singly charged and neutral neon atom.

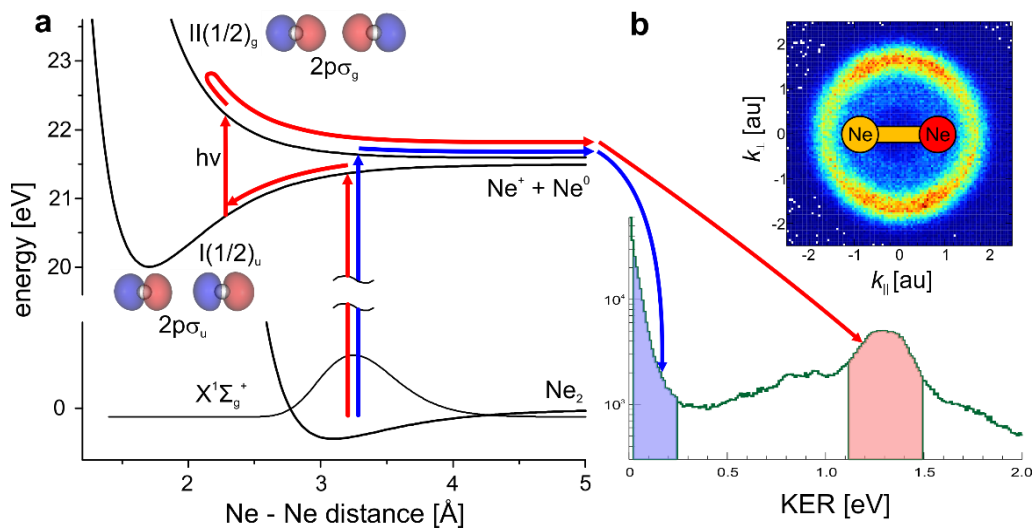

Fig. 1. a) Relevant potential curves of the neutral and singly charged neon dimer. Potential curves are taken from [22]. Spin-orbit splitting was simulated by separating asymptotic parts of the I and II curves by 100meV. b) Kinetic energy release (KER) for the $Ne^+$-$Ne^0$ dissociation channel. Two ionization pathways leading to $Ne^+$-$Ne^0$ dissociation are shown by red (indirect) and blue (direct) arrows. The inset shows the electron momentum distribution in the molecular frame, when both pathways are considered. The red side of the sketched

molecule defines the momentum direction of the detected neon ion. The corresponding orbitals of the neutral dimer are shown next to the potential curves.

The kinetic-energy release (KER) acquired by two neon counterparts after dissociation of the dimer shows two distinct structures around 0.25 eV and 1.3 eV (Fig.1b). The low-energy part corresponds to liberation of an electron from the $2p\sigma_g$ orbital, which results in direct dissociation of the dimer along its II(1/2)g ionic state (Fig. 1a, blue path). The KER around 1.3 eV is reached by absorption of one additional photon from the laser field after initial evolution along the I(1/2)u potential curve (removal of the electron from the $2p\sigma_u$ orbital). This transition lifts the dimer up to the II(1/2)g ionic state, but at a shorter internuclear distance resulting in gain of higher KER (Fig 1a, red path) (see also [23]).

For observing double-slit interference, it is decisive that the difference between initial phases of partial waves $\Delta\phi$ is the same for each event. For the optical case, this is trivially true, while for the homonuclear molecular case this is not always granted, since the electrons emerging from molecular orbitals of *gerade* and *ungerade* symmetries have $\Delta\phi$ which differ by π. This converts fringes to anti-fringes and washes away the interference pattern. It is the reason why almost all molecular double-slit experiments are performed on $H_2$ with only one symmetry (compare $N_2$ [7] and $Ne_2$ [13] for exceptions). As shown in Fig. 1 the dissociation process allows us to separate the two cases of emission from a *gerade* and *ungerade* orbital by selecting events from a certain region of KER. If no selection of the KER is made, implying both ionization channels are mixed, the photoelectron spectrum in the molecular frame shows no interferences (Fig 1b, inset). The separation is shown in Fig. 2. The electron removed from the *gerade* $2p\sigma_g$ orbital shows an interference pattern with the minimum along the $k_\perp$ axis at $k_\parallel=0$, whereas the ionization from the *ungerade* $2p\sigma_u$ orbital results in a maximum. This "swapped" interference behavior is understood given the p-character of the orbitals. The interference is seen indeed in the final electron momentum, which is acquired in a laser field after ionization rather than in the initial momentum at the tunnel exit.

Normalization of the electron spectra of the dimer by the single-ionization spectrum of a neon atom recorded in the same measurement allows us to remove an intrinsic ionization weighting, which has a "doughnut"-like shape due to tunneling ionization in a circular laser field. After such normalization, the interference strips perpendicular to the internuclear axis become clearly visible. These structures resemble the double-slit interferences described by equation (1) with $\Delta\phi$ being either 0 or π.

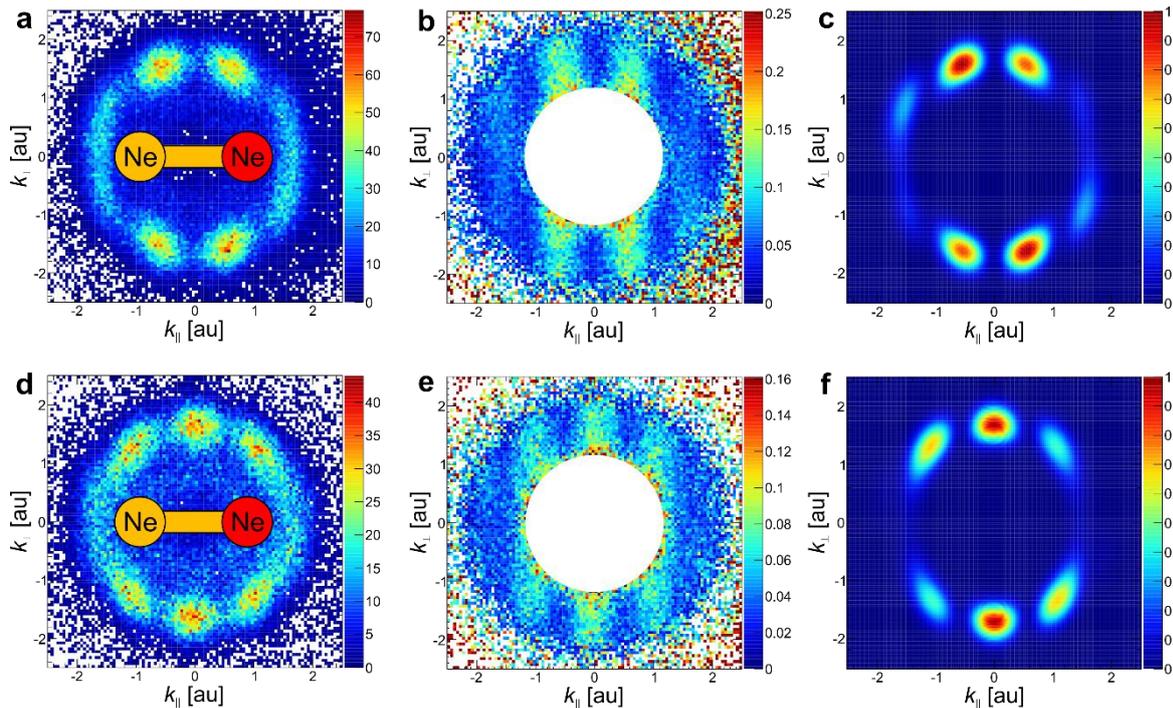

Fig. 2. Photoelectron momentum distributions in the molecular frame as defined in the method section: a) and d) - measured, b) and e) are the same as distributions in a) and d) but normalized to the monomer distribution in order to remove ionization weighting of the final momenta. c) and f) are simulated spectra with two coherently superimposed atomic distributions. Red side of the sketched molecule defines the momentum direction of the measured neon ion.

Theoretically, two atomic photoelectron momentum distributions were coherently superimposed in order to reproduce the double-slit interference pattern (fig.2 c,f). For each atomic spectrum, the TDSE was solved numerically within the single-active-electron approximation. An initial state was chosen to be a *p* orbital with lobes pointing along the hypothetical internuclear axis, thus, resembling "one half" of a 2*p*σ orbital of the dimer. The final momentum distributions were then projected onto the molecular frame (see Methods). This approach (fig. 2 c,f) accurately reproduces the double-slit interference, apart from the finite contrast in the fringes. This contrast, as the more complete theory reveals, is mainly due to interaction of the outgoing electron with the neighboring atom in the dimer (see Supplementary Information).

Pronounced interference patterns have also been observed upon ionization of the neon dimer in a linearly polarized laser field (Fig. 3). Here the dimer axis was postselected to lie within ±20 degrees to the field polarization direction. The two-center interference changes the photoelectron distribution significantly: for direct ionization of $Ne_2$ the pronounced minimum appears at zero momentum, where naturally, according to the tunneling theory, the maximum of the distribution resides.

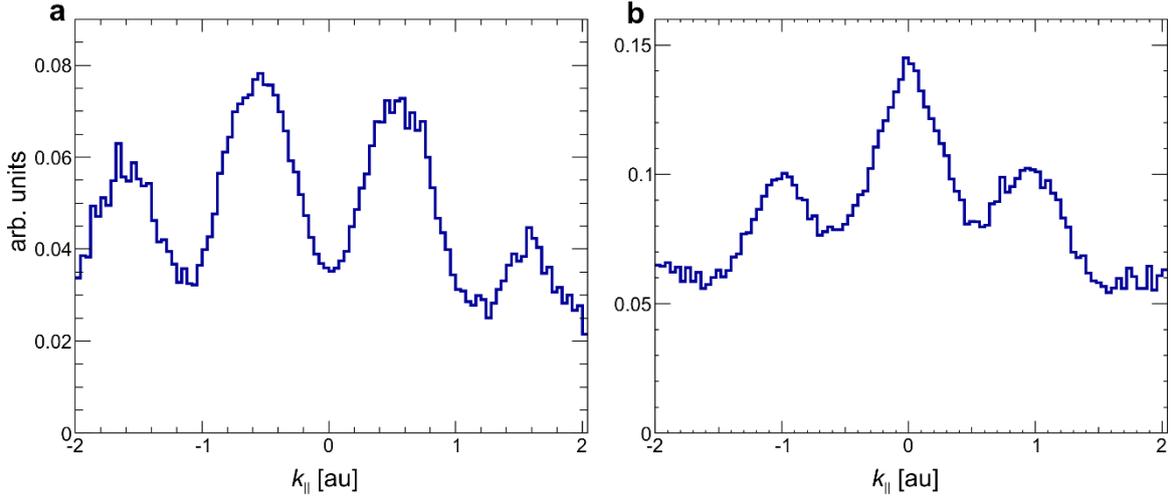

Fig. 3. Normalized photoelectron momentum distributions of Ne$_2$ projected to the molecular axis for the direct (a) and indirect (b) ionization pathways. The spectra were generated by selecting only the events, where the dimer axis lies within +/-20 degrees to the polarization direction of the laser field. Original spectra are divided by the corresponding spectrum of the monomer in order to remove the ionization weighting.

The dependence of the interference on the distance between the slits ($|\vec{R}|$) allows, in turn, to utilize the fringe pattern as a tool to measure the bond distance, which is the basis for diffractive imaging. In Fig. 4 we show the sensitivity of this approach. In addition to the interference, the ion momentum gives a second independent measure of the internuclear distance[6]. The II(1/2)g potential maps the ground-state density distribution of the neon dimer onto the final momentum (energy) of fragments[24]. Resolving the distance between two centers, one can see how the interference maxima move apart with decreasing separation between the two atomic centers (which corresponds to the higher ion momenta, Fig. 4). This observation is fully reproduced using equation (1) with $\Delta\phi=\pi$ (Fig. 4b). Here, the ion momentum was deduced by mapping internuclear distance to kinetic energy due to motion along the II(1/2)g potential.

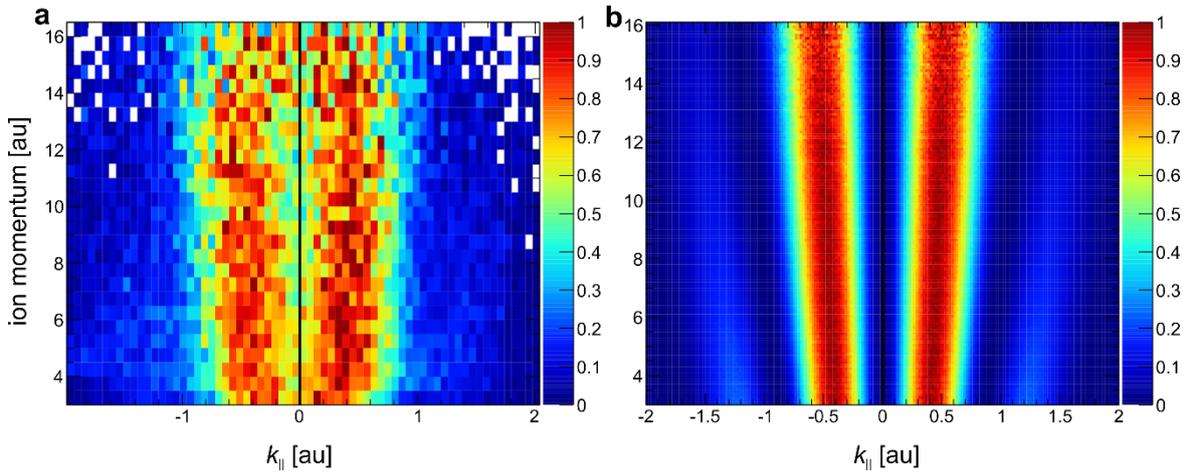

Fig. 4. Dependence of the two-center interference for the direct ionization pathway on the internuclear distance encoded in the momentum of the ion. a) experiment, b) classical simulation using ground state wave function and the II(1/2)g potential curve shown in Fig. 1. Each row is normalized to its maximum value in order to remove weighting caused by the ground state probability distribution. In the simulation the electron momentum distribution was weighted by a Gaussian distribution according to tunneling theory.

To conclude, we have observed two-center interference in the two-dimensional photoelectron momentum distribution in the molecular frame upon ionization of neon dimer by a strong laser field. By postselecting the dissociation channel of the dimer we were able to select the orbital the electron was liberated from upon tunneling ionization. This postselection allows for switching between *gerade* and *ungerade* interference patterns in the molecular frame photoelectron distribution, depending on the orbital symmetry. Turning it around, one might obtain the phase along the molecular orbital by considering the corresponding interference pattern. The finite contrast in the interference pattern has been attributed to interaction of the liberated electron with the neighboring neutral atom in the dimer. Extension of the "molecular" double-slit experiment to strong-field ionization allows for better understanding of molecular ionization and opens up new ways - ultimately time-resolved - of probing fundamental concepts of quantum mechanics.

**Acknowledgments**

The authors acknowledge financial support in the frame of the Priority Program "Quantum Dynamics in Tailored Intense Fields" (QUTIF) of the German Research Foundation.

**Methods**

*Experimental*

The Ne dimers were prepared in a molecular beam under supersonic expansion of gaseous neon at a temperature of 60 K through a 5 μm nozzle. The nozzle temperature was stabilized within ±0.1 K by a cryostat (Model 32B, Cryogenic Control Systems, Inc.) cooled with liquid helium. The optimum dimer yield was found at a nozzle back pressure of 3 bar. Neon dimers were selected from the molecular beam by means of matter wave diffraction using a transmission grating with a period of 100 nm. The selection allowed increasing the relative yield of $Ne_2$ to 20% with respect to the monomer.

The neon dimers were singly ionized by a strong ultra-short laser field (40 fs -FWHM in intensity -, 780 nm, 8 kHz, Dragon KMLabs). The field intensities were $7.3*10^{14}$ W/cm$^2$ (Keldysh parameter γ=0.72) in case of circular polarization and $1.2*10^{15}$ W/cm$^2$ (γ=0.4) in the experiment with linearly polarized light. The 3D momenta of the ion and electron after ionization were measured by cold target recoil ion momentum spectroscopy (COLTRIMS). In the COLTRIMS spectrometer a homogenous electric field of 16 V/cm for circularly polarized light, or 23 V/cm in case of linear polarized laser field, guided the ions onto a time- and position-sensitive micro-channel plate detector with hexagonal delay-line position readout [25] and an active area of 80 mm. In order to achieve 4π angle detection of electrons with momenta up to 2.5 au, a magnetic fields of 12.5 Gauss was applied within the COLTRIMS spectrometer in the experiment with a circularly polarized laser field. In the case of linearly polarized light a magnetic field of 9 Gauss was utilized. The ion and electron detectors were placed at 450 mm and 250 mm, respectively, away from the ionization region.

*Molecular frame photoelectron spectra*

The photoelectron momentum distributions with respect to the molecular axis shown in Fig. 2 were generated in the following way. Only ionization events have been considered, where ion and electron momentum vectors lie within slices along polarization plane, defined by conditions $|p_x|<0.55$ au for electrons and $|p_x|<3.5$ au and $|p_x|<12.0$ au for ions from direct and indirect dissociation channels, respectively (x-direction is the light propagation direction). These conditions ensure that the angle between a momentum vector and the polarization plane does not exceed 45° in the worst case. For the majority of events this angle is however smaller than 30°. Both, electron and ion, momentum vectors were projected onto the polarization plane. The projection of the ion momentum defines the $k_{||}$ direction, whereas two components, $k_{||}$ and $k_{\perp}$, of electron projection are plotted in Fig. 2. This type of molecular frame transformation avoids nodes along the dimer axis. It does not conserve the product $\vec{k} \cdot \vec{R}$, but the loss of contrast in the interference patterns is negligible. Another type of transformation, "natural", where the ion momentum vector, not its projection, defines the $k_{||}$ direction is presented in the Supplementary Information.

*Theory*

Starting from a 2*p* orbital aligned with the molecular axis, we solve the three-dimensional time-dependent Schrödinger equation (TDSE) in single-active-electron approximation with the split-operator method on a Cartesian grid with 512 points in each dimension, a grid spacing of 0.25 au and a time step of 0.02 au. While propagating up to a final time T=1500 au, outgoing parts of the wave function are projected onto Volkov states[26]. The potential for a single neon atom is chosen as in [27] but with the singularity removed using a pseudopotential for angular momentum $l = 1$ as described in [28]. The pulse has a 12-cycle sin² envelope and peak field strength of 0.096 au. To obtain the momentum distribution for the dimer we multiply two copies of the atomic distribution by $e^{\pm i\vec{k}\cdot\vec{R}/2}$ respectively (R/2=2.93au) and then add them coherently with an additional factor of $\pm 1$ depending on the type of interference, *gerade* or *ungerade*. To account for different possible orientations of the dimer with respect to the polarization plane, we vary the angle between them in 8 steps to cover a range from 0 to 45 degrees, project the molecular PMDs onto the polarization plane and add these projections together with their geometrical weights to obtain the final PMDs shown in Fig. 2.